# X-ray spectroscopy of Gamma-Ray Bursts

Luigi Piro[a*]

[a]*Istituto Astrofisica Spaziale e Fisica Cosmica, INAF, Roma 00133, Italy*

**Abstract**

In this review we briefly summarize the recent developments in the research on Gamma-Ray Bursts, and discuss in more details the recent results derived from X-ray spectroscopy, in particular the detection of X-ray narrow features and their implication on our understanding on the origin of GRB. Finally, we outline the importance of high resolution spectroscopy in X-rays, which can provide new clues on the nature of progenitors, and a powerful probe of the early Universe and primordial galaxy formation .

*Keywords:* Gamma-Ray Bursts; High energy Astrophysics; Cosmology; X-ray spectroscopy ;

PACS: 95.55.Ka; 95.75.Fg;  95.85.Nv; 97.60.Sm; 98.70.Rz

## 1. Introduction

Since 1996, the research on Gamma-Ray Burst has made a leap forward with the precise and fast locations catered for by the X-ray BeppoSAX mission [1,2]. Key milestones have been the discovery of long time scale emission, named afterglow, the distance-scale determination – GRB are taking place in distant galaxies -, and the likely association of these powerful explosions with the latest stages of very massive stars, suggested also by the association of few events with Supernovae [3]. We have also understood the basic mechanism producing the Gamma-ray burst and its associated afterglow in terms of a collimated highly relativistic expanding medium, the so called fireball model.

Recently, the field is focusing into the origin of the progenitor and on the mechanisms responsible for the production of the energy. At the same time, new classes of GRB phenomena have been discovered, the so called dark GRB[4] and X-ray flashes [5], whose origin is still to be comprehended, as it is also the case of short GRB.  Finally, it is becoming increasingly evident that the study of the early universe and its cosmological evolution can benefit of GRB research.  In this review we will focus on issues related to X-ray spectroscopic measurements and some scientific perspectives for  high resolution X-ray spectroscopy studies of GRB.

## 2. X-ray lines and progenitors

[*] Corresponding author. Tel.: +39-06-4993-4007; fax: +39-06-20660188; e-mail: piro@rm.iasf.cnr.it.

The GRB and its afterglow are very well explained by the fireball model [6], in which a highly relativistic outflow from the central source produces the observed emission. On the other hand, this process essentially loses "memory" of the central source: the shocks that are thought to produce the GRB and afterglow photons take place over a distance scale that is about 10 orders of magnitude greater than the size of the central source. In addition, this is almost independent of the details of the central source, depending primarily on basic parameters as the total energy, the collimation angle of the outflow (jet), the fraction of energy in relativistic electrons and magnetic fields and the density of the external medium.

A very effective method to gather information about the progenitor is to study the environment in its surrounding. Specifically, if the progenitor is a massive star, we would expect the GRB explosion to take place in a dense region, the star-forming site where the progenitor formed  This is because a massive star evolves very quickly (about a million years). On the contrary, if the GRB is the result of the coalescence of two neutron stars, the GRB would go off far away from the star formation site, because of the long time needed for the system to evolve and due to the kick-off velocity of the system following the Supernovae explosions that formed the two neutron stars. In the first case we would expect that the interaction of X-ray ionizing photons with the close medium could produce X-ray lines, while this should not be the case in the neutron star coalescence scenario.

The most common feature observed in X-ray astronomy in several classes of sources is the iron line. Most of the searches for features in GRB have been therefore concentrated at the energies of this element that, in the rest frame are: 6.4 to 6.9 keV for $K_\alpha$ lines from neutral to H-like ions, 9.3 keV corresponding to the recombination edge in emission from H-like ions and 7.1 keV, the energy of the absorption edge from neutral iron in absorption. So far there are 6 independent measurements of iron features from 4 different satellites (Fig. 1). For the four burst in which there is an independent redshift measurement from optical spectra, the emission features detected in the afterglow phase are consistent with highly ionized iron, while in one case there is evidence of a transient absorption edge from cold iron lasting about 12 s. during the main GRB pulse. This set of measurements can be explained altogether as follows. The line medium is external to the fireball region, as suggested by the presence of the absorption edge. In the early phase of the burst this medium is still to be completely ionized by the GRB photons, thus producing an absorption edge from neutral iron. As the ionization front reaches out the external border of the medium, this becomes completely ionized [12], thus explaining the disappearence of the absorption edge. On a time scale given by the recombination time, electrons start to recombine on ionized iron, thus producing the emission line and recombination edge observed in the afterglow phase.

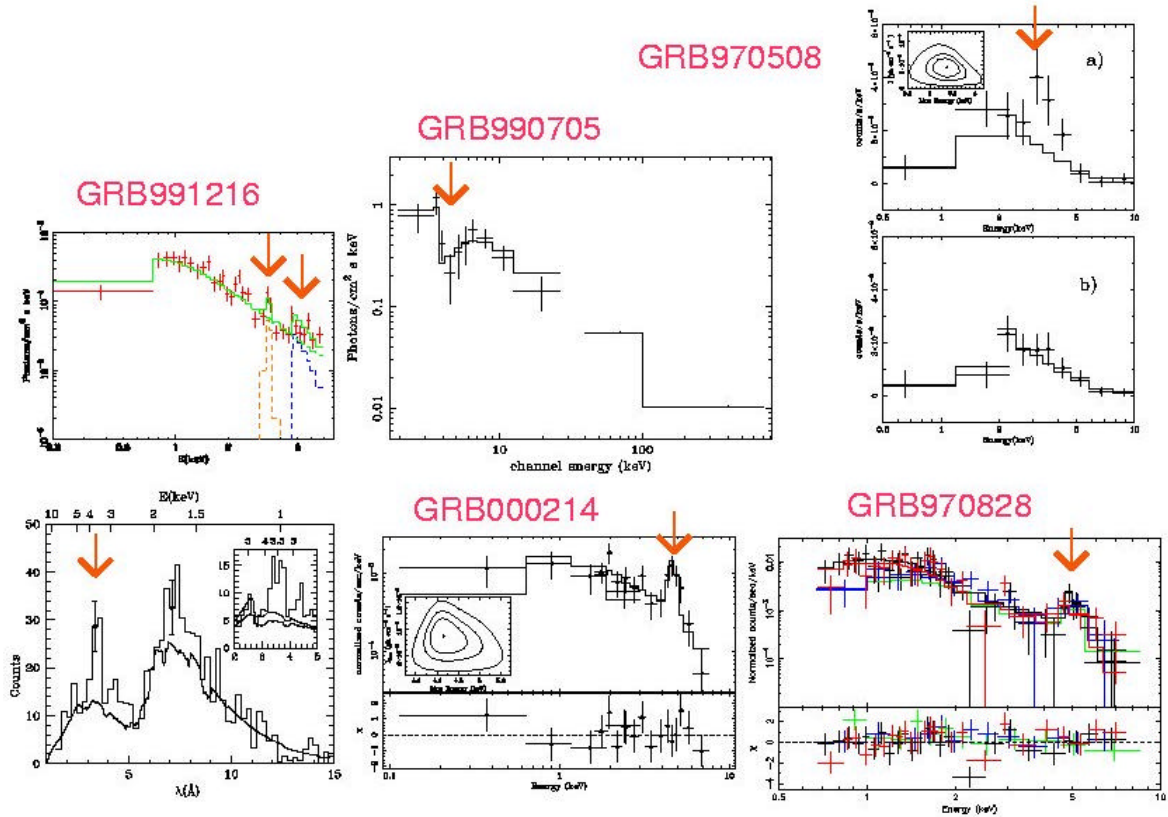

Fig. 1. Iron features observed in GRB X-ray spectra [7-11]

In an alternative model [13], it is assumed that the central source, after the event producing the GRB, continues its activity – at lower power -, heating and ionizing a close-by line emitting medium. The progenitor is likely a massive star that undergoes a core-collapse supernova explosion [14]. In the distant reprocessor scenario, this explosion takes place about a month before the event leading to the GRB [15], and are the Supernova ejecta illuminated by X-ray photons of the gamma-ray burst that produce the lines. This is also consistent with the line width observed in GRB991216 [10], that corresponds to an outflow velocity of 10% of the speed of light, as typically observed in Supernovae. In the local reprocessor scenario the two events are almost simultaneous.

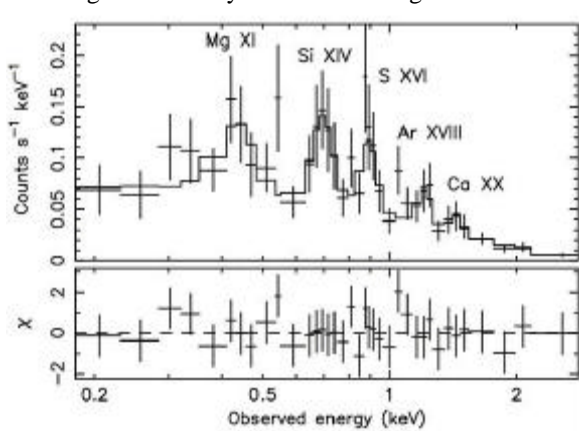

Fig. 2. Soft X-ray lines observed in the afterglow of GRB011211 [16]

In addition to Fe features, recent detections of soft X-ray lines by ionized elements as S, Si, Mg, (Fig.2) supports the association of GRB with SN-like explosions. In particular, those lines are blue-shifted with respect to the rest-frame energies by about 10% of the speed of light.

**3. X-ray spectroscopy of GRB and Cosmology**

*3.1. Finding primordial star-forming sites in the obscure Universe*

GRB are the most distant and brightest sources of the universe, with redshift ranging from 0.16 to 4.5. It is also likely that a substantial fraction of GRB lies at z>5 [17], but such events would be optically dark, because of the $Ly_\alpha$ absorption by hydrogen in the intragalactic medium (the so called $Ly_\alpha$ forest). Interestingly, there is a significant fraction of GRB that do not show an optical counterpart [4]. Since most of the redshift of GRB are derived from optical spectra, we are presently biased against high-z GRB. X-ray spectroscopy offers the mean (the other one being infrared measurements) to measure the redshifts of events at z>5. This is one of the primary motivations for pursuing high-resolution X-ray spectroscopy of GRB. Indeed, the most distant object observed so far is a quasar at redshift of about 7, an there is no information on the epochs between z=7 and 20, when the first population of stars and primordial galaxies formed. X-ray spectroscopy of GRB can thus provide a unique opportunity to localize and identify primordial star forming sites in the high-z universe.

*3.2. Tracing the evolution of metals in the Universe*

The X-ray spectra of several afterglows show evidence of substantial absorption in the range $10^{21}$-$10^{23}$ cm-2 [18]. This material is in the line of sight towards the GRB, and is likely connected with the gas in the star-forming site of the GRB. Present measurements are able to constrain the overall amount of column density produced by low-Z elements. By increasing the signal to noise and improving the resolution, it will become possible to measure the optical depth of each single element (basically: Si, S, Ar, Fe) thus building up the history of metal enrichment of the Universe from the local to the early epochs.

*3.3. Probing the Warm-Hot Intragalactic Medium*

An intriguing issue of Cosmology is the missing baryon mass in the local (z<2) universe. According to cosmological simulations [19] the intragalactic gas is accreting on dark matter structures, forming sheets and filaments that are heated to T $10^5$-$10^7$ °K. At this temperature the gas would then primarily emit and absorb photons in the soft X-ray range. Narrow redshifted resonant absorption lines from metals such as OVII in the spectrum of a bright background source [20] would provide an unambiguous signature of this medium. GRB are the best sources for this study (compared to quasars), being very bright even at z>2.

## 4. High-resolution X-ray spectroscopy experiments for GRB

For basically all of the measurements described above a high spectral resolution ($\Delta E$=1-4 eV) is needed, such as that afforded by X-ray microcalorimeters. For what regards the study of the prompt X-ray emission from GRB, we mention the IMBOSS experiment proposed for the ISS, described in more detail in this conference [21]. The large area (about 30m$^2$) afforded by XEUS [22], an ESA mission under study, coupled with the superb spectral resolution of TES microcalorimeters, will provide a spectrum of the needed quality even for an afterglow observed at about 1 day after the burst (Fig.3).

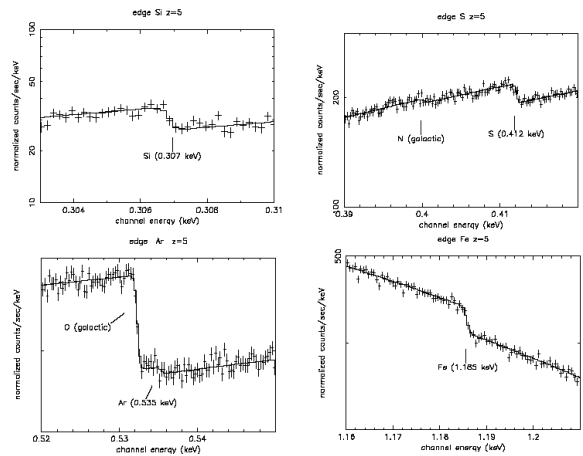

Fig. 3. Simulation of edges from Si, S, Ar and Fe in a 100 ksec observation by XEUS of an afterglow 1 day after the GRB, assuming z=5, $N_H$= 5x$10^{22}$ cm-2 and solar-like abundances.

Taking into account that the afterglow is extremely bright in the early phase (the flux decreases typically as $t^{-1.3}$), it will be possible to achieve similar results with a fast slewing satellite with a smaller X-ray telescope (but with a similar focal plane detector).